# Forces between Colloidal Particles in Aqueous Solutions Containing Monovalent and Multivalent Ions

Gregor Trefalt[1], Thomas Palberg[2], Michal Borkovec[1,*]

[1]Department of Inorganic and Analytical Chemistry, University of Geneva, Sciences II, 30 Quai Ernest-Ansermet, 1205 Geneva, Switzerland

[2]Institut für Physik, Johannes Gutenberg Universität Mainz, 55099 Mainz, Germany

*Corresponding author: Phone: ++41 22 3796405, Email: michal.borkovec@unige.ch

**Abstract**

The present article provides an overview of the recent progress in the direct force measurements between individual pairs of colloidal particles in aqueous salt solutions. Results obtained by two different techniques are being highlighted, namely with the atomic force microscope (AFM) and optical tweezers. One finds that the classical theory of Derjaguin, Landau, Verwey, and Overbeek (DLVO) represents an accurate description of the force profiles even in the presence of multivalent ions, typically down to distances of few nanometers. However, the corresponding Hamaker constants and diffuse layer potentials must be extracted from the force profiles. At low salt concentrations, double layer forces remain repulsive and may become long ranged. At short distances, additional short range non-DLVO interactions may become important. Such an interaction is particularly relevant in the presence of multivalent counterions.

**Keywords**

AFM, direct force measurement, colloidal probe technique, optical tweezers, DLVO theory, specific ion adsorption,

**Graphical Abstract**

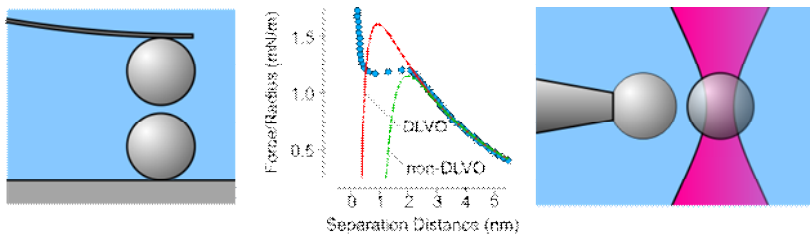

**Introduction**

Particle interactions govern a multitude of phenomena in colloidal suspensions. These include formation of fluid-like structures, colloidal crystals, gels, sediments, or growth of particle aggregates [1-6]. Here we focus on the relevant situation of aqueous dilute suspensions of identical (or very similar) particles, and address the case of interaction between particle pairs. The classical theory of Derjaguin, Landau, Verwey, and Overbeek (DLVO) stipulates that these interactions can be approximated by the superposition of two principal contributions, namely dispersion and double layer forces [1,2,7]. Attractive dispersion or van der Waals forces are always present provided the particles have different dielectric properties than the solvent. Repulsive double layer forces act between charged particles at sufficiently low salt concentrations. These forces may become very long ranged in deionized suspensions and induce liquid-like ordering or crystallization [3,8-10]. The less understood situation of forces acting between asymmetric particles or different types of particles will not be addressed here [11-13]. We will not discuss interactions in non-aqueous solvents as well [1,14].

Pair interactions govern the properties of dilute colloidal suspensions, but they may capture the behavior of relatively concentrated systems reasonably well too. In concentrated suspensions, however, additional types of interactions can be relevant. For repulsive inter-particle forces, additional conservative many-body interactions can be present, whereby the leading contribution is an attractive three-body potential [10]. For attractive inter-particle forces, the particles may stick to each other, often almost irreversibly [5,6]. Thereby, they form aggregates, which can even span the entire system and induce gelation [15]. The properties of such aggregated suspensions differ from their non-aggregated counterparts widely, especially concerning their structure and rheology. Finally, hydrodynamic interactions are induced through particle motion. Such motion may originate from thermal fluctuations, fluid flow, or presence of external fields [1]. However, these aspects are mostly relevant in concentrated suspensions, and they will not be discussed here in detail.

Recently, numerous authors have argued on theoretical grounds that DLVO theory may fail seriously [16-20]. One encounters three principal lines of argumentation. The first line argues that dispersion forces cannot be modeled by simple van der Waals contributions, but that an accurate description of these forces requires consideration of the surface structure and its roughness [17,21]. The second line queries the validity of the commonly used mean-field Poisson-Boltzmann (PB) approximation to treat the double layer forces, and calls for more detailed treatments, especially including finite ionic size, image charge effects, and ion-ion correlations [18,19,22]. Especially, the latter were claimed to be important in systems containing multivalent ions. Within the third line, one finally argues that other types of forces are equally relevant, such as hydration, depletion, or steric interactions [20,23,24].

For a long time, information about interactions between pairs of particles had to be deduced from structure factors, fluid-crystal phase boundaries, sedimentation profiles, or particle aggregation rates [1,3,5,8,9]. While much information could be obtained with these methods, the interaction forces could only be inferred indirectly. Only in the last two decades, experimental tools became available allowing direct measurements of interaction forces acting between colloidal particles [25-28]. The two most popular tools include the colloidal probe technique and optical tweezers. The colloidal probe technique is based on the atomic force microscope (AFM) [27-30]. Optical tweezers rely on an intense focused laser beam, which permits the trapping of a colloidal particle in its waist [25,26]. The position of the particles is then followed by video microscopy.

The present article reviews the status of recent force measurements between pairs of similar particles in aqueous suspensions. We will demonstrate that the newly available techniques make forces acting between individual colloidal particles routinely accessible. In contrast to some suggestions to the contrary, we will demonstrate that classical DLVO theory is capable to rationalize experimentally measured force profiles extremely well, even at low salt concentrations or in the presence of multivalent ions.

**Experimental Techniques**

Two main experimental techniques are currently available to measure forces between pairs of colloidal particles, namely the colloidal probe technique and optical tweezers. A scheme of these techniques is shown in Fig. 1.

The colloidal probe technique is based on the atomic force microscope (AFM) and has been originally developed to measure forces between a planar substrate and an individual colloidal particle, which was attached to the AFM cantilever [29-31]. Subsequently, the technique was extended to measurements between pairs of colloidal particles [27,28]. The second particle is then mounted to the substrate (see Fig. 1a). The particles can be attached by gluing [27,28], sintering [32,33], or spontaneous deposition on appropriately functionalized substrates [34,35]. The use of the latter technique is essential for measurements of forces between latex particles, as it avoids formation of nanobubbles on the hydrophobic particle surfaces. Such measurements require a lateral centering of the particles, which can be achieved by optical microscopy to sufficient accuracy. The forces are inferred from vertical approach-retraction cycles through deflection of the cantilever, which is measured by reflecting an incoherent laser beam from its back. The force is then obtained from the deflection by the Hooke's law, whereby the spring constant of the cantilever must be known. The latter quantity can be determined by various methods, for example, from the thermal fluctuation spectrum, by attaching small particles of known mass to its end, or by measuring its hydrodynamic drag [31,36]. The separation distance is obtained from the piezo-displacement, whereby the contact point is extracted from the constant compliance region, with a typical accuracy of one nm or better. The force resolution is limited by the thermal noise of the AFM cantilever, and typically is few tens of pN. However, the force resolution can be substantially improved by averaging subsequently recorded force curves. In this way, sub-pN force resolution can be achieved in favorable cases. When forces are strongly attractive, they cannot be probed with this technique due to the mechanical jump-in instability of the AFM cantilever. This problem can be circumvented by using a stiffer cantilever, but at the expense of force resolution.

The optical tweezers technique combines trapping of particles in an intense focused laser beam and the accurate tracking of their positions with video microscopy. The steepness of the harmonic trap potential depends on the focal laser intensity and can be calibrated from the Brownian excursions of the trapped particle [26]. Tracking algorithms rely on the intensity distribution in the recorded images [37]. The publicly available algorithm of Crocker and Grier [25] performs well for large inter-particle distances, but corrections must be introduced at closer distances [38]. Early work deduced the pair interaction potential from the trajectory of two particles released from two closely spaced optical traps [25]. Later, the potential was inferred from the time-averaged positions of two more softly trapped particles as a function of the distance from the trap [10,26]. Alternatively, one particle can be fixed with a micro-syringe and only one particle is optically trapped [39] (see Fig. 1b). The force resolution of these techniques is well in the sub-pN regime, but the accessible force range is more limited than for the colloidal probe technique. Measuring stronger attractive forces with the tweezers techniques is problematic, as distances between the surfaces of the particle should remain at least few nm, and particles should not stick to each other.

The principal advantage of both techniques is that a symmetric geometry is automatically realized. The symmetry can be further checked by repeating the measurements with different pairs of particles. A truly symmetric geometry is difficult to achieve with alternative direct force measurement techniques, with the exception of the surface forces apparatus (SFA) [40]. This difficulty is especially inherent to the sphere-plane geometry, which is frequently used with the colloidal probe AFM [29,30] or total internal reflection microscopy (TRIM) [41,42]. The disadvantage of all techniques available for the measurement of forces between pairs of individual colloidal particles is that one can routinely measure only relatively large particles, typically with diameters down to about 1 μm. This limitation mainly originates from the optical resolution of the microscope, which is used to manipulate the particles. However, given the substantial efforts to extend these techniques to smaller particle sizes, we expect that direct force measurements with substantially smaller particles will become possible in the near future.

**DLVO Theory**

Experimentally measured force curves are often interpreted within DLVO theory. Here, we provide a brief overview of this classical theory, and the reader is referred to textbooks for more details [1,2]. The basic assumption of this theory is that the force $F$ between two colloidal particles has two contributions, namely

$$F = F_{vdW} + F_{dl} \qquad (1)$$

where $F_{vdW}$ is the van der Waals or dispersion force, and $F_{dl}$ is the double layer force. There are various levels of approximation how these two contributions are being evaluated. Let us first discuss the dispersion forces, and double layer forces afterwards.

Dispersion forces can be simply estimated with the non-retarded van der Waals expression within the Derjaguin approximation [1]

$$\frac{F_{vdW}}{R} = -\frac{H}{12h^2} \qquad (2)$$

where $h$ is the smallest surface separation, $R$ the particle radius, and $H$ the Hamaker constant. Neglecting retardation effects is justified at short separations. The more accurate Lifshitz theory suggest that the dispersion force decays more quickly at larger distances, which is usually parameterized through a distance-dependent Hamaker constant [1]. The Derjaguin approximation stipulates that the surface separations in question are much smaller than the particle radius ($h = R$). Generally, the Derjaguin approximation is sufficient to interpret measured interparticle forces, since the particles in question are relatively large. For this reason, we report the experimentally measured forces as the ratio $F/R$.

Typically, the electrical double layer is normally treated on one of two levels of approximation, namely the non-linear Poisson-Boltzmann (PB) theory or its linearized version, the Debye-Hückel (DH) theory [1]. For sufficiently low electric potentials, these theories are equivalent. The easiest way to calculate double layer forces is by invoking the superposition principle for two charged plates within DH theory. The force acting between two particles is then obtained with the Derjaguin approximation, and one finds a simple exponential dependence [1]

$$\frac{F_{dl}}{R} = 2\pi\varepsilon_0\varepsilon\kappa\psi_{dl}^2\,e^{-\kappa h} \qquad (3)$$

where $\varepsilon_0$ is the permittivity of vacuum, $\varepsilon$ the dielectric constant of water, $\psi_{dl}$ the electric potential of the diffuse layer, and the screening parameter $\kappa$ is given by

$$\kappa^2 = \frac{2q^2 N_A I}{\varepsilon_0 \varepsilon kT} \qquad (4)$$

where $q$ is the elementary charge, $N_A$ is the Avogadro's number, $k$ the Boltzmann constant, $T$ the absolute temperature, and $I$ is the ionic strength. The latter quantity is defined by

$$I = \frac{1}{2} \sum_i z_i^2 c_i \qquad (5)$$

whereby the electrolyte solution contains ions of type $i$ with valence $z_i$ and molar concentration $c_i$. The Debye length is given by $\kappa^{-1}$ and it characterizes the range of the double layer force. In a solution of an ionic strength of 1 M it has a value of about 0.30 nm, and it increases with the inverse square root of the ionic strength. Within DH model, the diffuse layer potential $\psi_{dl}$ is related to the surface charge density $\sigma$ by a simple proportionality relationship [1]

$$\sigma = \varepsilon_0 \varepsilon \kappa \psi_{dl} \qquad (6)$$

At low salt concentrations, double layer forces can be long ranged, and the Derjaguin approximation may fail. However, the interaction potential energy can be expressed in a simple analytical form [1,43,44]

$$U_{dl}(h) = \frac{q^2 Z^2}{4\pi \varepsilon_0 \varepsilon} \cdot \left( \frac{e^{\kappa R}}{1+\kappa R} \right)^2 \cdot \frac{e^{-\kappa r}}{r} \qquad (7)$$

where $Z$ is the particle charge in units of $q$ and $r$ the center-to-center distance, which his given by

$$r = 2R + h \qquad (8)$$

This expression is also referred to as the screened Coulomb or Yukawa potential. In salt-free suspensions, the Debye length is calculated from the counterions concentration only [43,44]. The force acting between particles can be obtained from the derivative, namely $F_{dl} = -dU_{dl}/dr$. For small separations $h \approx R$ one recovers eq. (3) by invoking eq. (6).

The DH superposition approximation inherent to eqs. (3) and (7) fails under various circumstances. Most importantly, the diffuse layer potential may not be sufficiently small in magnitude to ensure the validity of the DH approximation, as the threshold is about 25 mV in monovalent systems. However, these expressions remain valid at sufficiently large distances, provided the diffuse layer potential $\psi_{dl}$ or the charge $Z$ is replaced by the effective quantities $\psi_{eff}$ or $Z_{eff}$. When the PB theory is used, one can demonstrate that the effective quantities saturate for highly charged surfaces [3,9,45]. In monovalent salt solutions and for the planar geometry, the saturation limit is [46]

$$\psi_{\text{eff}} \to \frac{4kT}{q} \; ; \; 103 \text{ mV} \qquad (9)$$

where the numerical value refers to room temperature. In spherical geometry, one has [47]

$$Z_{\text{eff}} \to \frac{R}{l_B} \cdot (4\kappa R + 6) \qquad (10)$$

where $l_B = q^2 / (4\pi\varepsilon_0 \varepsilon kT)$ is the Bjerrum length, which is about 0.71 nm at room temperature. For large particles ($\kappa R \gg 1$), these expressions become again equivalent by virtue of eq. (6). For small particles, the dependence becomes linear in the radius with a pronounced transition region observed at very low salt concentrations [46,48].

At smaller separations, charge regulation effects can be important. These effects arise, since the surface charge may vary (regulate) upon approach through adsorption or desorption of ions. Some surfaces may maintain a constant charge (CC) upon approach, but the magnitude of the surface potential will then increase. In other situations, the magnitude of the charge density decreases, while the surface maintains a constant potential (CP). In real systems, the situation is often intermediate, and is related to the details of the adsorption isotherm. The simplest way to capture this dependence is the constant regulation (CR) approximation [13].This approximation introduces an additional regulation parameter $p$, which assumes the values of $p = 1$ for CC and $p = 0$ for CP. The superposition approximation given in eq. (3) corresponds to the DH model with $p = 1/2$. For special adsorption isotherms, extreme charge regulation effects are possible, and they have been referred to as sub-CP behavior ($p < 0$) [49]. As will be discussed below, many of these effects can be successfully treated within the PB model. In this case, the respective force profiles are best obtained numerically. Further details on charge regulation phenomena can be found elsewhere [13].

**Forces at High Salt Concentrations**

In concentrated salt solutions, the double layer forces are fully screened and the particle interactions are dominated by dispersion forces. Measurements of such forces are shown in Fig. 2. The salt concentrations were chosen to be sufficiently high to fully eliminate the contribution from double layer forces, which is normally the case for concentrations exceeding 500 mM. Consider first the force measurements between polystyrene latex particles (Fig. 2a) [35]. The measured force profile can be relatively well described with the non-retarded expression given in eq. (2). By fitting the profile, one obtains a Hamaker constant of $3.5\times10^{-21}$ J. Note that this value is substantially smaller than the available theoretical estimates, which are around $1.3\times10^{-20}$ J [50]. One might suspect that retardation effects and screening by the salt might be responsible for this reduction. However, these effects are relatively unimportant. According to Lifshitz calculations, retardation effects become considerable only for distances beyond 10 nm. While screening effects are non-negligible at all distances, they lead only to a reduction of about 30% in the relevant distance range. In our view, the much more substantial decrease of the Hamaker constant observed experimentally is due to roughness effects. The latex particles have a root mean square (RMS) roughness of about 0.8 nm. Lifshitz calculations of dispersion forces including surface roughness have demonstrated that a comparable roughness may explain a similar reduction of the apparent Hamaker constant [21]. Earlier estimates of the Hamaker constant of latex particles, which are available from aggregation rate measurements, lead to substantially larger values. However, these estimates seem to be influenced by the presence of additional attractive short-range non-DLVO forces [51]. However, these short-ranged forces cannot be easily measured with the AFM at high salt concentrations due to the mechanical instability of the cantilever.

The importance of surface roughness can be equally illustrated with silica particles, see Fig. 2b [33,52]. The data can be again well described with eq. (2). The two data sets represent measurements carried out with the same particles, but these particles were heated at two different temperatures. Forces acting between the particles heated at 1150°C can be modeled with a Hamaker constant $2.0\times10^{-21}$ J. This value is well comparable to the one measured with the AFM in the sphere-plane geometry [53] and to the currently best available estimate of $1.6\times10^{-21}$ J [54]. On the other hand, forces for the particles heated at 1050°C are compatible with a substantially smaller Hamaker constant around $3.0\times10^{-22}$ J. We suspect that this major difference in the strength of dispersion forces is again caused by differences in surface roughness. The RMS roughness of the particles heated at 1150°C is 0.8 nm, while the ones at 1050°C is 2.5 nm [52]. The substantial variation in the strength of dispersion forces acting between silica surfaces reported experimentally [33,52,53,55] is probably caused by variations in roughness of the different particles used. These effects are currently being addressed theoretically as well [17,21].

The data for the silica particles shown in Fig. 2b further illustrate the presence of a repulsive non-DLVO force. This short-ranged force $F_{sr}$ can be modeled with an exponential profile, namely

$$\frac{F_{sr}}{R} = Ae^{-bh} \qquad (11)$$

where $A$ is the amplitude of the interaction and $b^{-1}$ its range. To interpret the data, this force was added to the dispersion force profile, which was shifted by 0.60 nm. The range of this force, which is compatible with the data is $b^{-1} = 0.3$ nm. Since this force is repulsive, the amplitude $A$ is positive. Similar short-ranged forces acting between silica surfaces with a similar range were observed with the colloidal probe technique in the sphere-plate geometry [56,57]. This force either represents a hydration force or originates from the presence of hairy layer of polysilicilic acids tails on the particle surface [23,24].

Only few force measurements have been reported for concentrated salt solutions containing different types of ions [21,35,51]. Nevertheless, the available data suggests that in such media dispersion forces always dominate and that these forces are independent of the type of ions present.

**Forces at Lower Concentrations of Monovalent Salts**

The characteristic feature of double layer forces is the decrease of their range with the salt concentration, and in many situations, their exponential distance dependence. These features are illustrated in Fig. 3a by direct force measurements between silica particles in KCl electrolyte [33]. In the semi-logarithmic representation used, one observes almost straight lines, especially at larger distances. The slope of these lines reflects the inverse Debye length $\kappa$, which can be reliably calculated with eq. (4) from the known salt concentration. The solid lines shown in Fig. 2a are obtained with a numerical solution of the PB model within the constant regulation approximation with a regulation parameter $p = 0.58$. The intercept is related to the diffuse layer potential. For the force profile measured at 1 mM, the resulting diffuse layer potential is $\psi_{dl} = -61$ mV. At larger separations, the force profile can be equally well interpreted with the DH model, but in this case an effective potential of $\psi_{eff} = -52$ mV must be used. Note that the magnitude of the effective potential is indeed bounded by the saturation value given in eq. (9).

The diffuse layer potential calculated with the PB model is shown in Fig. 3b. Its magnitude decreases with increasing salt concentration, and can be well described with the charge-potential relationship for a charge density of $-6.5$ mC/m$^2$. The respective DH calculation based on eq. (6) is shown for comparison. Incidentally, the sign of the diffuse layer potential cannot be inferred from the measured force profiles, since they depend on the square on the potential, see eq. (3). This sign must be obtained by other means, for example, from the chemical identity of the surface groups.

At shorter distances, additional effects have to be considered [35]. Charge regulation is the most important one. These effects are illustrated by the force profiles acting between sulfate latex particles in 10 mM KCl solution shown in Fig. 3b. The experimental data are well described with the PB model within the CR approximation. This model uses a diffuse layer potential of $\psi_{dl} = -47$ mV and a regulation parameter $p = 0.41$. The corresponding results for the CC ($p = 1$) and CP ($p = 0$) are also shown. While the CR approximation captures most features of the double layer force accurately, a current problem of the CR model concerns the salt dependence of the regulation parameter. While adsorption models predict a strong increase of this parameter with increasing salt concentration, the available experiments rather suggest that this parameter remains approximately constant [13].

Experimental data shown in Fig. 4a (arrow) illustrate the mechanical jump-in instability. While the measured force profile should ideally follow the arrow, whose slope reflects the spring constant of the cantilever, the actual trace deviates due to the resulting acceleration of the cantilever and the occurring hydrodynamic interactions [35]. Nevertheless, the onset of the jump-in should be correctly reflected by the calculated force profile. When attractive dispersion forces are included in the calculation, the jump-in is predicted to be located at too small distances. This discrepancy indicates the presence of an additional attractive non-DLVO force. This short-ranged force can be modeled with the exponential law given in eq. (11). Since the force is attractive, the amplitude $A$ is negative. The data are again compatible with a range of $b^{-1} = 0.3$ nm. When this additional force is included in the model, the position of the jump-in can be described accurately. Such short-ranged non-DLVO forces have been shown to be essential to properly describe measured force profiles acting between latex particles, especially to rationalize the correct jump-in position [35].

At lower salt concentrations, double layer forces acting between colloidal particles can become very long ranged. Due to their weakness, such force profiles have been measured with optical tweezers. Forces acting between latex particles at moderately low salt concentrations are shown in Fig. 5a. While the data extend to lower forces than typically accessible with the colloidal probe, the overall behavior is very similar. As expected, the force profiles decay exponentially, whereby their range is given by the Debye length, see eq. (3). In such situations, force profiles measured with optical tweezers have been show to agree well with colloidal probe measurements [58].

Very long ranged interaction energy profiles can be observed in deionized suspensions. Such suspensions have to be prepared with mixed-bed ionic exchangers, whereby one can reach concentrations <1 µM in closed cells and <10 µM in suspension in contact with ambient air. In these situations, the ionic concentration in the solution can be dominated by the counterions originating from the suspended particles [59]. In the ideal case, such systems are close to salt-free. Nevertheless, the interaction potentials follow the same exponential dependence, as illustrated in Fig. 5b. Due to the long-ranged nature of these interactions, one must use the screened Coulomb potential given in eq. (7) since the Derjaguin approximation used in eq. (3) fails. At the lowest ionic strength investigated, the resulting Debye length is 185 nm and the effective charge 15'900. For these conditions, eq. (10) predicts the saturation charge of 23'800. In such systems, the effective charges can also be estimated in the fluid-phase from osmotic pressure and structure factors measurements [9,60,61] or from conductivity [62]. Effective charges are further accessible in colloidal crystals from the shear modulus or coexistence lines [63-65]. However, the presence of many-body forces in such systems complicates the concept of the effective charge substantially [63,66] and different effective charges may be obtained from different experiments on the same systems [67,68].

The appropriateness of the screened Coulomb potential in suspensions with extremely low ionic strengths was also established in confined two-dimensional slits by inverting the measured pair correlation functions recorded by video microscopy [44]. While this approximation is valid at low particle concentrations, additional attractive component in the interaction potential became apparent with increasing particle concentrations. This effect is being referred to as macroion shielding, and has also been confirmed theoretically with combined PB and Brownian dynamics simulations in charged particle suspensions [66]. This picture is further consistent with the presence of attractive three-body forces, which were quantified by optical tweezer measurements [10] and numerical studies [69].

**Forces at Lower Concentrations of Multivalent Salts**

The theoretical community devoted substantial attention to forces between particles or surfaces in the presence of multivalent ions [18,19,22]. Recently, one could witness an important complementary development, namely that such force profiles are becoming available experimentally to good precision [34,35,39]. Distinguishing the role of multivalent counterions and coions is crucial. The valence of the counterion has an opposite sign than the charge of the particles, while for the coions the respective sign is the same. Moreover, soluble salts containing multivalent ions are normally combined with monovalent ions, and the resulting electrolytes become asymmetric. The consideration of this aspect is essential, when force profiles are calculated with the PB model.

Let us first discuss the situation of multivalent counterions. The first notable point is that force profiles between particles in solutions containing divalent (and sometimes trivalent) counterions can be still very well described by DLVO theory. Such observations were made with negatively charged particles in the presence of divalent and trivalent cations [33,39] as well as positively charged particles in the presence of divalent anions [34]. Nevertheless, there are two important differences with respect to the monovalent situation. First, the multivalent ions are more effective to induce screening, since their valence is weighted more strongly in the ionic strength. At equal concentrations, multivalent ions induce a substantially smaller Debye length. Second, diffuse layer potentials are lower in magnitude than in the monovalent case. The latter difference is likely caused by specific adsorption of the multivalent counterions to the particle surfaces, which leads to partial charge neutralization, and lowers the magnitude of the surface potential. This adsorption process is the principal reason why the DH theory actually is – somewhat counter intuitively – a better approximation in the presence of multivalent counterions than in the monovalent setting. Similar observations were made in the sphere-plate geometry by TIRM [70].

We have already mentioned that multivalent counterions adsorb to oppositely charged surfaces, and thereby they reduce the magnitude of diffuse layer potential. With increasing valence, however, this adsorption process becomes more prominent, and can lead to complete charge neutralization and subsequent charge reversal. As an example, consider forces between positively charged amidine latex particles in the presence of tetravalent anions $Fe(CN)_6^{4-}$. The force profiles at low concentrations are shown in Fig. 6a and at higher concentrations in Fig. 6b. The resulting diffuse layer potentials are given in Fig. 6c. At low salt concentrations, the forces are repulsive at larger distances, but they become attractive at shorter distances, mainly due to dispersion forces. As the salt concentration increases, the double layer repulsion weakens, and disappears completely at 60 µM. At this concentration, the surface charge is neutralized and the double layer potential vanishes. When the concentration is increased further, the diffuse layer repulsion sets in anew, and its range decreases due to increased screening. Such reentrant double layer repulsion is typical for charge reversal processes. The magnitude of the diffuse layer potentials remains low, and therefore the DH approximation is again reasonable for this system. Analogous charge reversal processes have been revealed by direct force measurements for sulfate latex particles and multivalent polyamine cations [35]. The presence of such charge reversal processes were observed by electrokinetic techniques earlier [71,72]. The diffuse layer potentials are effective potentials describing the ionic profiles further away from the surface, and they can be even different in sign than the bare surface charge of the particles [73,74].

Multivalent counterions of sufficiently high valence further induce characteristic short-ranged attractive non-DLVO forces. The presence of such forces is evident when comparing the two calculated attractive force profiles shown in Fig. 6a. The more weakly attractive one corresponds to the dispersion force alone, while the more strongly attractive one includes an additional exponential force given by eq. (11). Since the force is attractive, the amplitude $A$ is negative. The range of this force turns out to be $b^{-1} = 1.8$ nm, thus substantially larger than the short-ranged forces discussed above. The magnitude of the amplitude increases with concentration and sharply decreases at higher salt concentration. This non-DLVO force is absent in concentrated salt solutions. Similar short-range forces were reported between sulfate particles in the presence of cationic polyamines [35] and between silica particles in the presence of $La^{3+}$ [33]. A similar additional attractive force was measured in the sphere-plate geometry in the presence of trivalent cations [75,76]. These observations suggest that short-ranged non-DLVO attractions are induced by multivalent counterions of sufficiently high valence. The origin of these forces is currently not clear to us, but they could be related to ion-ion correlations, charge fluctuations, surface charge heterogeneities, or depletion forces [18-20,22,77,78].

While we have argued that the DH approximation can be reasonable in the presence of multivalent counterions, this argument does not apply to multivalent coions. As such ions will in general not adsorb to like-charged surfaces, the magnitude of the diffuse layer potential will remain high. This scenario is illustrated in Fig. 7. The two subfigures illustrate the analogous situations for different concentrations both signs of charge, namely for positively charged particles and multivalent anions, and for negatively charged particles and multivalent cations [79]. The counterions are always monovalent. One observes that the force profiles are clearly non-exponential, and this character becomes increasingly pronounced with decreasing salt concentration.

The observed force profile can be well described with PB theory, provided the equation is solved for the appropriate asymmetric electrolyte. The reason for this non-exponential behavior is that the multivalent coions are expelled from the surface, and in this region the surface charge is screened by counterions only. Such salt-free regime leads to a non-exponential concentration profile, and thus induces the corresponding decay in the force profile. Only at relatively large distances, one recovers the exponential DH decay, whereby the Debye length given by the appropriate ionic strength. The corresponding effective potentials can become relatively large. For example, for the data measured in 0.1 mM solution of $K_4Fe(CN)_6$, the effective potential $\psi_{eff} = -176$ mV. The magnitude of the corresponding saturation potential is 316 mV, which is substantially larger than the value for a monovalent electrolyte given in eq. (9) [79,80].

**Conclusion**

This article reviews the current status of direct force measurements between individual colloidal particles in aqueous solutions containing simple salts. Such measurements can be currently best achieved with the colloidal probe technique based on the atomic force microscope (AFM) or optical tweezers combined with video microscopy. The major advantage of both techniques is that one can realize almost perfectly symmetric systems. A specific advantage of the AFM is the possibility to investigate forces acting between anisotropic particles [11,12] or to combine various types of particles in a single experiment [13]. On the other hand, the tweezers techniques currently offer the best force resolution [39,58,81]. The disadvantage of both techniques currently is that one can investigate only relatively large particles, typically >1 µm in diameter. However, due to rapid technical developments, especially improved force resolution of the AFMs and higher acquisition rates in video microscopy, similar measurements should become possible with smaller particles in the near future.

In spite of these limitations, these techniques have already provided a wealth of detailed information concerning forces acting between colloidal particles in solution. One comforting finding is that the classical DLVO theory, especially when complemented with numerical solutions of the full PB equation including charge regulation, provides an astonishingly accurate description of the observed force profiles down to few nm, even in the presence of multivalent ions. However, dispersion forces are typically weaker than the ones predicted by Lifshitz theory for smooth bodies, and we suspect that surface roughness that is inherent to most colloidal particles is responsible for this weakening. Moreover, diffuse layer potentials or effective charges may depend on the salt concentrations in a non-trivial fashion and they have to be extracted from the measured force profiles. These quantities seem not easily accessible otherwise. While electrokinetic data may provide reasonable estimates of diffuse layer potentials in some situations, they reveal important discrepancies in others [21,35,82]. The origin of these discrepancies is currently difficult to pinpoint. The situation appears even more complicated in deionized suspensions, as effective potentials or charges vary with the particle concentration due to macroion shielding effects [44,63,66].

Direct force measurements with the AFM further reveal the existence of additional non-DLVO forces. The range of these forces is often just a few Angstroms, and they cannot be easily resolved with the currently available techniques. Such short-ranged forces are probably due to hydrophobic or hydration interactions [23]. In the presence of multivalent counterions, however, the non-DLVO forces are attractive and more easily accessible, since their range is about 1 nm [33-35,75]. It was not yet possible to pinpoint the origin of these forces, but they could be related to ion-ion correlations, charge fluctuations, surface charge heterogeneities, or depletion forces [18-20,22,77,78]. Future experimental and theoretical studies should help to clarify their origin.

**Acknowledgements**

This research was supported by the Swiss National Science Foundation (Project No. 159874 and 150631), University of Geneva, and Deutsche Forschungsgemeinschaft (DFG Grant No. Pa459/18 and Pa459/19). We acknowledge useful discussions with David Andelman, Klemen Bohinc, Marco Heinen, Christian Holm, Christophe Labbez, Leo Lue, Rudi Podgornik, and Martin Trulsson.

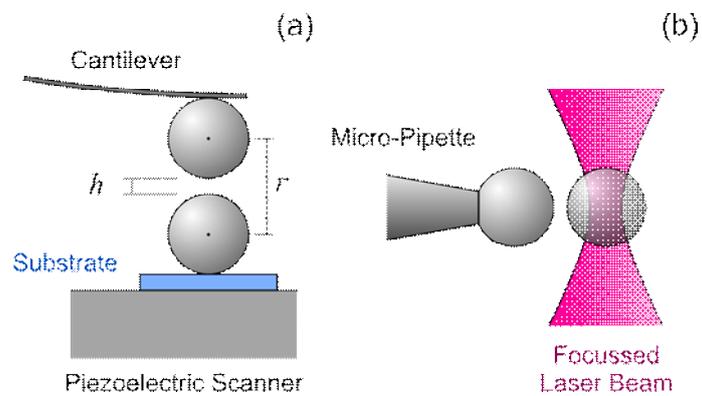

**Figure 1.** Scheme of the two main techniques in use to measure forces between pairs of individual colloidal particles. The closest surface separation distance *h* and the center-to-center distance *r* are also indicated. (a) Colloidal probe technique and (b) optical tweezers.

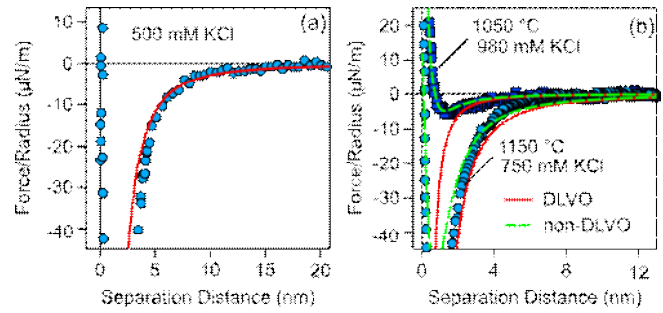

**Figure 2.** Dispersion forces between colloical particles measured with the AFM in concentrated KCl solutions adjusted to pH 4.0. The solid lines represent the dispersion force calculated with eq. (2). Data for (a) sulfate latex particles of 3.0 μm in diameter with $H = 3.5 \times 10^{-21}$ J [35] and (b) silica particles of 4.4 μm in diameter [35]. Prior to the measurements, the silica particles were heated at the temperatures indicated [33,52]. The fitted Hamaker constants are $H = 3.0 \times 10^{-22}$ J for heat-treatment at 1050°C and $H = 2.0 \times 10^{-21}$ J for 1150°C. The non-DLVO potential is also indicated whereby the dispersion force was shifted by 0.60 nm, the range of exponential force was $b^{-1} = 0.3$ nm. The amplitudes $A$ = 0.48 mN/m at 1050°C and $A$ = 0.16 mN/m at 1150°C.

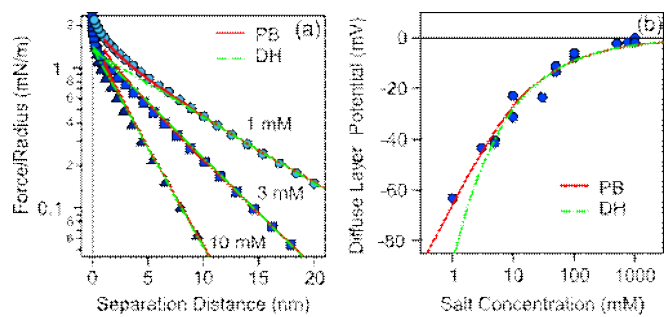

**Figure 3.** Double layer forces between silica particles heated to 1150°C measured with the AFM in KCl solutions of different concentrations adjusted to pH 4.0 [33]. Calculations based on the PB model with the regulation parameter of $p$ = 0.58 are compared with the DH superposition approximation. (a) Force profiles. (b) Diffuse layer potentials versus the salt concentrations. The lines are calculations of PB and DH theory with a surface charge density of –6.5 mC/m$^2$.

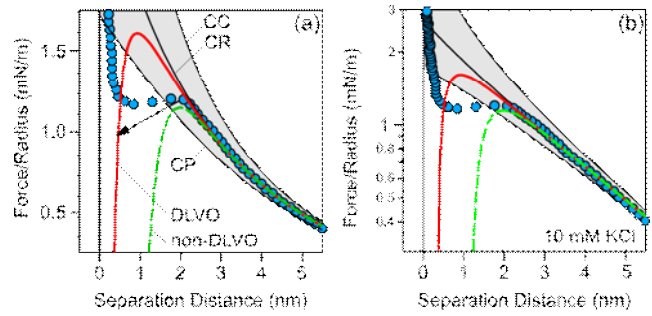

**Figure 4.** Forces between sulfate latex particles of 3.0 μm in diameter measured with the AFM in 10 mM KCl solution adjusted to pH 4.0 [35]. Comparison of the experimental data with DLVO theory based on the PB model and constant regulation approximation (CR) and including a non-DLVO attractive exponential force. The measured force profile shows jump-in instability (arrow). The grey region is bounded by the PB results for constant charge (CC) and constant potential (CP) boundary conditions. The parameters are $\psi_{dl}$ = –47 mV, $p$ = 0.41, $b^{-1}$ = 0.3 nm and $A$ = –70 mN/m. (a) Linear and (b) semi-logarithmic representation.

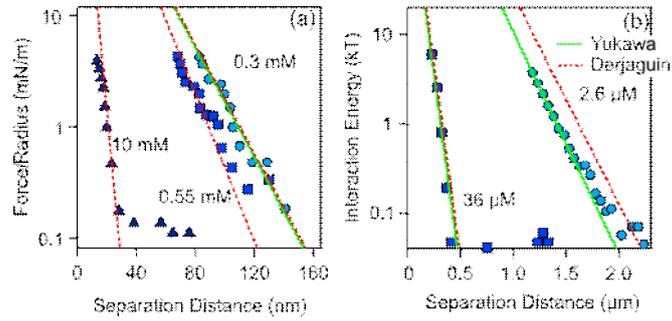

**Figure 5.** Interactions between charged particles measured with optical tweezers techniques. Screened Coulomb potential given in eq. (7) is compared with the Derjaguin approximation given in eq. (3). (a) Forces between polystyrene sulfate latex particles of diameter 2.3 μm in NaCl salt solutions [39]. The solid lines are calculated with DH theory with the Derjaguin approximation with the effective potential of –45, –52, and –26 mV for salt concentration 0.3, 0.55, and 10 mM. (b) Interaction potentials between silica particles of 1.5 μm in diameter at the ionic strength indicated [81]. The effective charges of 15'900 and 18'700 are used at ionic strengths 2.6 μM and 36 μM, respectively.

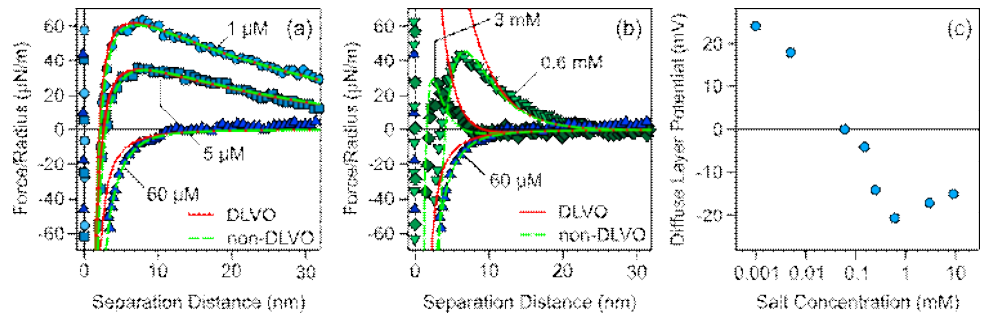

*(This is a two column figure. Do not print this text.)*

**Figure 6.** Behavior amidine latex particles of 1.0 μm in diameter in $K_4Fe(CN)_6$ salt solutions of concentrations indicated [34]. Comparison between measured and calculated force profiles (a) at lower and (b) higher salt concentrations. The calculations compare DLVO theory with the same resulting including an additional non-DLVO attraction. The parameters are $p = 0.50$, $H = 4.0 \times 10^{-21}$ J, and $b^{-1} = 1.8$ nm. (c) Fitted diffuse layer potentials versus the salt concentration.

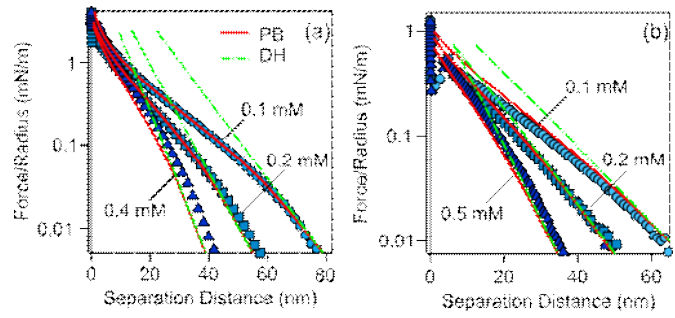

**Figure 7.** Forces acting between colloidal particles measured in the presence of multivalent coions. Experimental data are compared with calculations based on the PB model. The limiting DH law is also indicated. (a) Sulfate latex particles of 3.0 μm in diameter in $K_4Fe(CN)_6$ salt solution [79]. The parameters used are $\sigma = -9.3$ mC/m$^2$ and $p = 0.64$. (b) Amidine latex particles of 1.0 μm in diameter in $LaCl_3$ salt solution [74]. The parameters used are $\sigma = -3.9$ mC/m$^2$ and $p = 0.41$.